# *In vivo* compaction dynamics of bacterial DNA:
# A fingerprint of DNA/RNA demixing ?

Marc Joyeux[(#)]

*Université Grenoble Alpes, LIPHY, F-38000 Grenoble, France*

*CNRS, LIPHY, F-38000 Grenoble, France*

**Abstract:** The volume occupied by unconstrained bacterial DNA in physiological saline solutions exceeds 1000 times the volume of the cell. Still, it is confined to a well defined region of the cell called the nucleoid, which occupies only a fraction of the cell volume. This is puzzling, because bacterial DNA is not delimited by a membrane, in sharp contrast with the nucleus of eukaryotic cells. There is still no general agreement on the mechanism leading to the compaction of the DNA and the formation of the nucleoid. However, advances in *in vivo* sub-wavelength resolution microscopy techniques have recently allowed the observation of the nucleoid at an unprecedented level of detail. In particular, these observations show that the compaction of the nucleoid is not static but is instead a highly dynamic feature, which depends on several factors, like the richness of the nutrient, the cell cycle stage, temperature, the action of an osmotic shock or antibiotics, *etc*. After a short description of the electrolyte content of the cytosol and a brief overview of the different mechanisms that may lead to the formation of the nucleoid, this paper reviews some of the most fascinating recent results of *in vivo* sub-wavelength resolution microscopy. It is furthermore argued that these observations provide converging indications in favor of a model that describes the cytosol as an aqueous electrolyte solution containing several macromolecular species, where demixing and segregative phase separation occur between DNA and RNA (essentially rRNA and mRNA involved in translation complexes, but also the large amounts of rRNA synthesized at the *rrn* operons of cells growing in rich media). It is also pointed out that crowding may play a crucial role through its synergy with electrostatic forces. By constraining macromolecules to remain at short distances from one another and feel electrostatic interactions in spite of the strong screening exerted by electrolyte species, crowding favors stronger DNA/RNA demixing and nucleoid compaction.

**Keywords:**   Bacterial nucleoid, DNA compaction, segregative phase separation, demixing, sub-wavelength resolution microscopy

[(#)] email : marc.joyeux@univ-grenoble-alpes.fr

## 1. Introduction

Theory predicts that an unconstrained circular (closed) polymer with contour length $L = 1.5$ mm and persistence length $\xi = 50$ nm forms a coil with radius of gyration $R_g = \sqrt{L\xi/6} \approx 3.5$ µm [1], with loops that extend radially over distances significantly larger than $R_g$. This formula is derived under the assumption that the only contribution to the potential energy of the polymer stems from bending deformations and does not consider eventual electrostatic repulsion forces between the segments of the polymer, which would result in still larger values of $R_g$ and wider radial extension of the polymer coil. These values of contour and persistence lengths are those of the genomic DNA molecule of *Escherichia coli* bacteria, which is constrained to fit in a cylindrical cell with typical diameter 1 µm, length 2 µm, and volume 1 µm$^3$ in the stationary phase. The DNA molecule must therefore be compacted by about 3 orders of magnitude (in volume) to fit inside the cell [2]. It has been estimated, both theoretically [3] and from micro-piston experiments [4], that the free energy that must be supplied to compact the DNA from its unconstrained coil conformation to its dimensions inside the cell is of the order of 0.01 to 0.1 $k_B T$ per base pair, that is about $10^5$ $k_B T$ for the whole molecule. Moreover, the genetic material of prokaryotes is not enclosed in a membrane-bound nucleus, as is the case for eukaryotes. As a consequence, one would expect that the DNA molecule occupies the whole available space when constrained to fit in the cell. It has, however, been known for decades that this is usually not the case and that in most circumstances the DNA molecule occupies only a rather well defined fraction of the cellular volume called the *nucleoid*. According to standard estimations, the dry mass of the nucleoid is composed of about 60% to 80% of genomic DNA (the rest consisting of RNA and proteins) and the nucleoid occupies approximately 25% of the cell volume. In contrast, the cytosol outside the nucleoid is essentially devoid of DNA, except for short plasmids, and composed essentially of RNA and proteins. Owing to the lack of a bounding membrane, this localization of the genomic DNA molecule inside the nucleoid has puzzled the scientists since its very discovery about half a century ago and is still the subject of ongoing debates.

Progresses in this domain have been hindered for a long time by the small dimensions of bacteria, which are of the same order of magnitude as the resolution of traditional optical microscopes, and the fact that electronic microscopy observations depend dramatically on the procedure used to prepare the cells. Fortunately, the development of optical microscopy techniques with sub-wavelength resolution and of single-molecule imaging [5] has more recently allowed *in vivo* observations of nucleoids at an unprecedented level of detail. These fascinating new results show that the compaction of the nucleoid is not static but is instead a highly dynamic feature, which depends on several factors, like the richness of the nutrient, the growth rate, temperature, the action of an eventual osmotic shock, *etc*. For example, Fig. 1 shows images of cells growing in a rich medium and dividing every 30 min (left column) and of cells growing in a nutrient-poor medium and dividing approximately every 110 min (right column). Comparison of the two columns indicates that during rapid growth nucleoids are compact structures with well-defined shapes, while the edges of the nucleoids of cells growing slowly in a minimal medium are much more poorly defined and the nucleoids themselves are less compact [6-10]. Moreover, for given growth conditions, the compaction of the nucleoid fluctuates during the cell cycle. For example, Fig. 2 shows the organization of *E. coli* nucleoids at different cell cycle stages. It is seen in this figure that nucleoids are substantially more compact during the segregation of daughter chromosomes (cell with length 3.5 µm) than during replication (cell with length 2.7 µm) or after they have reached their home position (cell with length 3.9 µm) [11]. A third striking illustration is provided by the observation of cells treated with antibiotics that inhibit translation or transcription. As may be checked in Fig. 3(A) and Fig. 4(A), inhibition of translation by chloramphenicol leads to very compact nucleoids, while inhibition of transcription (and consequently also of translation) by rifampicin leads to expanded nucleoids that extend almost through the whole cell [8-10,12-16].

The goal of the present paper is to review some of the most exciting recent results of *in vivo* sub-wavelength resolution microscopy and to emphasize that they probably allow for a great step forward towards deciphering the principal forces involved in the dynamics of the bacterial nucleoid. More precisely, it will be argued that they provide converging indications that segregative phase separation induced by electrostatic repulsion between DNA and RNA (essentially ribosomal RNA) may be the leading force in the nucleoid compaction dynamics.

The paper is organized as follows. Section 2 provides a short description of the ionic content of the cell. It will be pointed out that among the many dissolved cationic and anionic species that compose the cytosol and are responsible for its peculiar properties, DNA and RNA molecules represent the only large ensembles of connected charges with uniform sign. A brief overview of the mechanisms that may lead to the formation of the nucleoid is next proposed in Section 3, leading to the conclusion that segregative phase separation induced by electrostatic repulsion between DNA and other negatively charged molecules may play an important role in the formation of the nucleoid. Section 4 consequently examines the possibility that compaction of the nucleoid may result from demixing between DNA and RNA and confronts this hypothesis with the results of high-resolution *in vivo* microscopy experiments. This is the key section of the paper. It will be argued that this interpretation is indeed supported by the observations reported above, as well as a couple of additional ones. Based on simulations performed with a coarse-grained model, Section 5 finally highlights the fact that crowding of the cytosol may be a mandatory ingredient to compensate for the short range of electrostatic interactions inside the cytosol. We conclude this review in Section 6.

## 2. The cytosol, a complex electrolyte solution

The cytosol of bacterial cells is an aqueous solution ($\approx$50-70% of water) that contains essentially three groups of molecules and ions, namely macromolecules (proteins and nucleic acids), small molecules (including precursors of macromolecules, metabolites and vitamins), and several inorganic ions and cofactors. Macromolecules account for nearly all of the $\approx$0.4 pg dry weight of the cytosol, with approximately $\approx$0.3 pg of proteins, corresponding to a concentration of amino-acids around 3 M, and $\approx$0.1 pg of nucleic acids, corresponding to a concentration of nucleotides around 300 mM [17]. DNA represents only about 10% of the content in nucleic acids of the cytosol (around 30 mM) [17], the major contribution (around 75%) arising from ribosomal RNA (rRNA), while transfer RNA (tRNA) and messenger RNA (mRNA) contribute more or less to the same extent as DNA. Moreover, while the *E. coli* genome is predicted to code for more than 4000 proteins, the most abundant ones in the cytosol are those involved in protein synthesis, most notably ribosomal proteins [18] ($\approx$20% of the cytosol protein content).

The concentration at neutral pH of charged species in the cytosol depends very sensitively on the growth medium and the stage of the cell cycle. Still, one can estimate that negative charges arise essentially from

- the phosphate groups of nucleic acids (300 mM) [17],
- the negatively charged residues of proteins, aspartic and glutamic acids (300 mM),
- the glutamate metabolites $Glu^-$ (20-150 mM) [17,19,20],

while positive charges arise principally from

- potassium ions $K^+$ (150-500 mM) [17,19,21,22], of which approximately 50% are bound to negatively charged molecules and 50% are free,

- the positively charged residues of proteins, arginine and lysine (300 mM),

- magnesium ions $Mg^{2+}$ (100mM) [22], nearly all of which are bound to nucleic acids,

- putrescine$^{2+}$ ions (5-35 mM) and spermidine$^{3+}$ ions (5-10 mM) [23].

It is particularly striking that DNA and RNA molecules represent the only large ensembles of connected charges with uniform sign. Indeed, the surfaces of most proteins display only small positively or negatively charged patches and the net charge of a protein is small compared to the total number of residues. Moreover, charged small molecules and ions are usually free to diffuse in the cytosol and/or bind to other molecules and ions that often carry the opposite charge. In contrast, the phosphate groups of DNA and RNA molecules are bound to other phosphate groups that carry the same negative charge, so that the positions (as well as the displacements) of these negatives charges are necessarily highly correlated.

This remark holds in particular for rRNA and ribosomes. Functional 70S ribosomes are composed of two subunits, the small 30S subunit, which consists of one RNA molecule (≈1500 nucleotides) and about 21 proteins, and the large 50S subunit, which consists of two RNA molecules (≈120 nucleotides and ≈2900 nucleotides, respectively) and about 31 proteins for *E. coli* cells. Each ribosome contains therefore approximately 4500 nucleotides and 7000 amino acid residues. Ribosomes have a diameter of about 20-25 nm and account for approximately 30% of the dry mass of the cell. Most ribosomal proteins are positively charged to keep the rRNA molecules together, but the total charge of ribosomal proteins is small compared to the total charge of rRNA. Indeed, a rough counting of charged residues indicates that the net positive charge carried by the proteins of one 70S ribosome is of the order of +460*e*, which must be compared to the -4500*e* negative charge carried by the rRNA molecules. As a first approximation, ribosomes may consequently be considered as spherical complexes with an almost homogeneous density of negative charges.

The overall picture that emerges from this description is therefore one of negatively charged macromolecules (DNA, RNA, and ribosomes) moving in an aqueous solution crowded with roughly neutral proteins and large amounts of positively charged ions or small molecules that are free to diffuse or bind to other species under the influence of electrostatic forces.

As will be discussed in more detail in the next Section, of crucial importance is the proportion of DNA and RNA phosphate charges that are actually neutralized by positive ions or small molecules, since this ratio is one of the key parameters for nucleoid compaction. Indeed, a substantial population of cations condenses into a small volume surrounding the (locally) cylindrical DNA and RNA polyanions and neutralizes part of their charges. The proportion of condensing cations depends sensitively on the valency of the cations, but rather little on their nature and overall concentration. Moreover, these cations must be viewed as roaming in the close neighborhood of the surface of the polyanions rather than being bound to particular sites. Manning's theory for counterion condensation [24] predicts that the neutralization ratio is $1 - b/(Z\ell_B)$, where $1/b$ is the density of charges along the polyanion (in units of the elementary charge $e$), $Z$ the valency of the cations, and $\ell_B = e^2/(4\pi\varepsilon_0\varepsilon_r k_B T)$ the Bjerrum length of the cytosol, that is, the separation at which the electrostatic interaction between two elementary charges $e$ is equal to thermal energy $k_B T$. Since the Bjerrum length of water at 25°C is of the order of $\ell_B \approx 0.7$ nm and the inverse of the charge density of bare DNA of the order of $b \approx 0.17$ nm, the neutralization ratio is approximately 76%, 88%, 92%, and 94% for cations with valency $Z$=1, 2, 3, and 4, respectively. This implies than even if the cytosol would contain only monovalent cations, then already only one fourth of the charges along the DNA would be able to contribute to electrostatic repulsion between different DNA duplexes. As described above, the cytosol however contains cations with different valences, which compete for counterion condensation, and the neutralization ratio is usually larger than 76%. Formulae for estimating the total fraction of neutralized DNA phosphate charges when the buffer contains two cationic species with different valences can be found for example in Ref. [3].

Small ions that do not participate in the neutralization of the charges of the macromolecules contribute to the properties of the electrolyte solution by shortening the Debye length $\lambda_D$, which sets the scale for the variations in the electric potential and the concentrations of charged species. Since $\lambda_D = \sqrt{(\varepsilon_0\varepsilon_r k_B T)/(2N_A e^2 I)}$, where $N_A$ is the Avogadro number and $I = \frac{1}{2}\sum_i c_i z_i^2$ the ionic strength of the solution containing species $i$ with molar concentration $c_i$ (in mole/m$^3$) and charge $z_i$ (in units of the elementary charge $e$), the Debye length is close to 3.1 nm for an aqueous solution at 25°C containing a monovalent salt at 0.01 M concentration, but reduces to approximately 1.0 nm at 0.1 M salt concentration, thus implying that a tenfold increase in salt concentration decreases the action range of electrostatic forces by a factor larger than 3.

## 3. Overview of the mechanisms that may lead to the formation of the nucleoid

As already mentioned, the unconstrained genomic DNA molecule of *E. coli* cells is expected to form a coil with volume about 3 orders of magnitude larger than that of the cell because of the stiffness of the strands and the electrostatic repulsion between the phosphate groups. Still, when constrained to fit in the cell, together with large amounts of RNA (about 10 times the mass of the DNA) and proteins (about 30 times the mass of the DNA), plus large amounts of (essentially) positive ions and small molecules, the DNA molecule does not spread homogeneously throughout the cell but localizes instead in the nucleoid. The mechanisms that may lead to such a separation between a region rich in DNA and a region depleted in DNA have intrigued the scientists for the past decades. They have been reviewed recently [25] and will consequently be overviewed here only briefly, for the sake of clearer discussions in the following sections.

*DNA charge neutralization and fluctuation correlation forces induced by polycations.* Studies of solutions containing long DNA molecules and various concentrations of salts with different valences have shown that DNA compacts strongly when 90% of the phosphate charges are neutralized (estimated according to Manning's theory [24]), irrespective of the exact nature of the salts [26], provided that the solution contains a significant portion of trivalent or higher valency cations (see above). Very compact globules are similarly obtained by adding polyamines with three or four positive charges to very dilute DNA solutions [27]. The proposed explanation is that correlations between the fluctuations of the ionic clouds condensed around each duplex lead to an attractive force between like-charged DNA duplexes, which is strong enough to compete with stiffness and electrostatic repulsion if the valency of the counterions is sufficiently large [28].

*Macromolecular crowding and depletion forces.* In as crowded an environment as the cytosol, attractive forces can however arise even between chemically and electrically non-interacting species. Indeed, each macromolecule is surrounded by an excluded volume, which other macromolecules cannot penetrate. When two macromolecules come close to each other, their excluded volumes overlap and more volume is consequently made available for the other molecules in the solution, thereby increasing the entropy of the system and opposing the separation of the two neighboring macromolecules [29]. The resulting entropic attraction force is called a depletion force and is known to provoke the collapse of long DNA molecules upon increase of salt and simple neutral polymer concentrations above certain thresholds [30], large amounts of (eventually monovalent) salts being needed to screen as efficiently as possible all electrostatic repulsion forces.

Both mechanisms described above lead, however, to DNA structures with nearly liquid crystalline ordering, which are much denser than the nucleoid of live bacteria. These are moreover essentially all-or-none processes, with the DNA being either in the coil state or the condensed one, but not in-between. Old *in vitro* results have been confirmed by recent *in vivo* experiments, where NaCl was dissolved rapidly into cultures of *E. coli* cells growing in a rich medium and their hyper-osmotic response was monitored by fluorescence microscopy [12]. Images of the nucleoids show that they shrink rapidly (in less than 10 min) down to very compact globules, thereby expelling the RNA polymerase (RNAP) enzymes that were located inside before the osmotic shock. Since the compaction of genomic DNA inside the bacterial nucleoid is much milder than the condensation provoked by these two mechanisms (and moreover fluctuates with cell cycle), it is probable that fluctuation correlation forces and depletion forces are not the leading compaction mechanisms in living cells. Still, the neutralization and screening of DNA charges by multivalent and monovalent cations certainly represent crucial contributions to the compaction of the macromolecule by reducing substantially the required free energy.

*Supercoiling*. Supercoiling is often quoted as one of the principal mechanisms for bacterial DNA compaction, although most experimental and theoretical results suggest that its role is actually rather limited. For example, relaxation of supercoiling through inhibition of the gyrase activity results in a rather limited increase in the size of *E. coli* nucleoids [12,31] and the theoretical estimation of the radius of gyration of the supercoiled genomic DNA of *E. coli* leads to a value not smaller than 3 μm [32].

*Nucleoid Associated Proteins*. The nucleoid contains tens of thousands of proteins, essentially Nucleoid Associated Proteins (NAPs), RNA polymerase, DNA polymerases, and many species of the transcription factor, the concentrations of which vary widely with cell cycle. Fis, Hfq, and HU are the most abundant NAPs in growing cells, while Dps and IHF predominate in the stationary phase [33]. Some of the NAPs have so-called architectural properties, in the sense that they can bridge (H-NS), bend (IHF, HU, Fis) or wrap (Dps) the DNA, and it is often suggested that these architectural properties play an important role in the compaction of the nucleoid. Indeed, several NAPs, including Fis [34] and IHF [35], induce gradual and strong DNA compaction *in vitro*, albeit at concentrations much larger than the intracellular ones. Other NAPs induce local compaction. For example, complexes of DNA and HU [36] or LrpC [37] proteins form nucleosome-like beads, while sufficiently large concentrations of H-NS lead to globular DNA/H-NS complexes [38], which reorganize into zipped duplexes upon deposition onto mica surfaces [39,40]. *In vivo* overproduction of H-NS moreover leads to too compact nucleoids and is lethal [41], while bacterial cells lacking both HU and Fis display instead large decondensed nucleoids [42]. There is, however, no physical reason why bending proteins should be able to compact DNA on large scales and standard intracellular concentrations of bridging proteins are not sufficient to achieve this goal [25]. Consequently, while it is probable that bridging proteins play a role in the partitioning of the nucleoid into macro-domains [43,44], a level of organization that will not be discussed in the present review, their architectural properties are most likely not responsible for the overall compaction of DNA inside the nucleoid.

Up to now, we have consequently ruled out the hypothesis that mechanisms based on DNA charge neutralization and screening (fluctuation correlation forces and depletion forces) and mechanisms relying on DNA's and NAPs' specific properties (DNA supercoiling and NAPs' architectural binding modes) may lead the compaction of bacterial genomic DNA inside the nucleoid. The remaining possibility is that direct interactions between the DNA molecule and other macromolecules in the cytosol, involving either hydrogen bonds or electrostatic forces between non-neutralized charges, play this role. If such interactions are attractive, then one may observe associative phase separation (also called complex coacervation), with one phase rich in both DNA and the other molecule (the nucleoid) and another phase depleted in both of them (the rest of the cytosol). If these interactions are instead repulsive, one may under certain conditions observe segregative phase separation, with one phase rich in DNA and depleted in the other molecule (the nucleoid), and a second phase rich in the other molecule and depleted in DNA (the rest of the cytosol). For segregative phase separation to take place, the effective interaction between the different molecular species must be repulsive. Application of Flory-Huggins polymer solution theory to solutions containing two polymer species A and B [45], indicates that such an effective repulsion between A and B requires that the $\chi_{A/B}$ parameter be positive, where $\chi_{A/B}$ compares the strength of the pair interaction between A and B segments with the average of the pair interaction between two A segments and the pair interaction of two B segments. If all these pair interactions are of the same magnitude (so that $\chi_{A/B} \approx 0$), there is no selectivity and no segregation occurs.

*Associative phase separation*. It has been known for several decades that long cationic poly-amino acids, like poly-L-lysine, induce a gradual compaction of DNA when the number of amino acid residues is increased with respect to the number of DNA phosphate groups [46,47], compaction reaching its maximum when the concentrations of the two oppositely charged groups become equal [47]. This makes polylysines an interesting vector for the delivery of therapeutic DNA. Moreover, in eukaryotic cells, highly positively charged histone proteins package and order the DNA molecule into nucleosome subunits, which are further compacted into the chromatin fiber. The point, however, is that neither long polycations nor free proteins with large positive charges are known to be present at sufficiently large concentrations in the cytosol, so that complex coacervation probably does not play an important role in the compaction of bacterial DNA.

*Segregative phase separation*. The hypothesis that the formation of the nucleoid may result from segregative phase separation has received little attention up to quite recently. The first demonstration of segregative phase separation involving DNA was probably provided by the investigation of salt solutions containing from 5 to 15% (w/v) of BSA (bovine serum albumin) proteins [48,49], which are globular macromolecules with a net negative charge of approximately -18$e$ distributed almost homogeneously all over their surface. Compaction of DNA is progressive and increases with BSA content. At 5% BSA, elongated coils and compacted globules coexist along the same DNA molecule, while at 15% BSA the DNA is so tightly compacted that it merely appears as a bright spot in fluorescence microscopy [49]. It was also checked that increasing the monovalent salt concentration from 150mM up to 200 mM at 15% BSA provokes the decompaction of the DNA molecule, which returns back to a coiled conformation. This indicates that electrostatic forces play an important role in the compaction mechanism and that it is crucial that the non-neutralized electrostatic charges along the DNA not be screened too severely for compaction to occur. However, it may be felt that these results are not fully conclusive, since attractive interactions between DNA and BSA proteins and the formation of relatively weak BSA-DNA complexes and coacervates have been reported [50]. Therefore, it cannot be completely excluded that the fewer positively charged patches on the surface of BSA molecules are sufficient to let these proteins bind to DNA and cause conformational changes and compaction. Still, it has been shown even more recently that DNA compaction also takes place in solutions containing a few percents of negatively charged silica nanoparticles with diameters ranging from 20 to 135 nm [51], which unambiguously confirms that electrostatic repulsion between DNA and negatively charged particles can indeed lead to phase segregation and DNA compaction. Although it is only scarcely mentioned in the list of possible mechanisms, segregative phase separation may therefore strongly contribute to, and eventually lead, the formation of the nucleoid in bacterial cells. Interestingly, simulations performed with a coarse-grained model also support this hypothesis [25].

The bacterial cytosol does not contain large concentrations of BSA proteins and no silica nanoparticles at all, but it does instead contain large amounts of negatively charged RNA molecules, so that the question that naturally arises from the preceding discussions is whether demixing between DNA and RNA may play an important role in the compaction of the bacterial genomic DNA. This question is addressed in the following section.

## 4. Segregative phase separation, the key player for nucleoid formation ?

Owing to the conclusions of Sections 2 and 3, it is tempting to hypothesize that segregative phase separation between negatively charged macromolecules, that is DNA and RNA, plays a key role in the formation of the nucleoid. Recent *in vivo* high-resolution microscopy results, which will be reviewed in this section, support this hypothesis.

*Functional ribosomes are excluded from the nucleoid*. As mentioned in Section 2, ribosomes are the most negatively and uniformly charged macromolecular entities in the cytosol besides DNA. Functional 70S ribosomes have a diameter of about 20-25 nm, a net charge (before neutralization) of about -4000$e$, and account for approximately 30% of the dry mass of the cell. Functional ribosomes are moreover bound to mRNA molecules, which contribute an additional 1000 to 1500 negative fundamental charges to the translation complex. If segregative phase separation induced by electrostatic repulsion is to play a role in the formation of the nucleoid, then it may be expected that the strongest demixing mechanism involves DNA and functional ribosomes. Quite interestingly, recent *in vivo* microscopy experiments confirm the longstanding knowledge that most functional ribosomes are excluded from the *E. coli* nucleoid [14,15,52]. Such a mutual exclusion of DNA and ribosomes [14,15,52] may in turn be interpreted as the fingerprint of DNA/ribosomes demixing and of the segregative phase separation leading to the formation of the a phase rich in DNA (the nucleoid) and a phase rich in ribosomes (the rest of the cytosol). Stated in other words, the $\chi_{DNA/70S}$ parameter (see Section 3) is positive and sufficiently large to induce demixing.

*Ribosomal subunits are not excluded from the nucleoid*. It has been observed very recently that, while functional 70S ribosomes are excluded from the nucleoid, this is not the case for free 30S and 50S subunits, which are able to diffuse inside the nucleoid [52]. Assembly of functional ribosomes on nascent or mature mRNA can therefore take place throughout the nucleoid and translation can start immediately, which is consistent with the observation that for some genes the overall transcription/elongation rate is controlled by the translation rate [53]. The low number of functional ribosomes observed in the nucleoid however indicates that translation complexes are rapidly expelled from the nucleoid when the RNA is not (no longer) bound to the transcribed gene. These observations may be interpreted as an indication that $\chi_{DNA/30S}$ and $\chi_{DNA/50S}$ are small compared to $\chi_{DNA/70S}$, so that the first phase separation that occurs upon decrease of the volume of the solvent corresponds to the demixing of DNA and full 70S ribosomes, with free 30S and 50S subunits being distributed more or less equally between the two phases. It is tempting to ascribe the smaller values of $\chi_{DNA/30S}$ and $\chi_{DNA/50S}$ compared to $\chi_{DNA/70S}$ to the smaller volumes and charges of 30S and 50S subunits compared to full 70S ribosomes, but the relation between the $\chi_{A/B}$ parameter and the physical or geometric properties of the A and B species is not straightforward and the interactions between each of these species and the solvent may also play a role.

*The nucleoid of cells grown in poor media is less compact*. As illustrated in Fig. 1, a striking result of recent *in vivo* microscopy experiments is the observation that the nucleoids of cells growing rapidly in rich media are compact structures with well defined shapes, while the nucleoids of cells growing slowly in minimal media have a less condensed organization and less defined edges [6-10]. Let us first examine the case of nutrient-poor media. It is known that cells growing in poor media express a large part of their ≈4000 genes at low to moderate levels [54,55]. Most of the 1500-5000 RNAP enzymes contained in the cell consequently diffuse homogeneously in the nucleoid in search for promoter targets, spending 85% of the time bound non-specifically to DNA [10]. After finding and fixing to their promoters, the RNAP enzymes use an incoming flux of ribonucleotides to synthesize mRNA molecules (≈1000-1500 nucleotides). The homogenous distribution of bound RNAP molecules is clearly seen in the top plot of Fig. 4 of Ref. [10] and the left column of Fig. 5, which shows 3D surface renderings of observed RNAP and DNA distributions in a cell grown in a minimal medium. Diffusing 30S and 50S ribosomal subunits may eventually assemble on some of these nascent or mature mRNA, for which translation starts immediately. When not bound to a nascent mRNA, the translation complexes are rapidly expelled from the nucleoid towards the rest of the cytosol, where translation continues. Since in nutrient-poor media many genes are expressed at a low to moderate level, the transcription/translation process ensures that there is a continuous and homogenous assembly of large negatively charged translation complexes throughout the nucleoid. Stated in other words, in nutrient-poor media the nucleoid is permanently out of equilibrium, with segregation forces tending to compact the nucleoid and the continuous and homogeneous synthesis of translation complexes inside the nucleoid tending to oppose compaction.

*In rich media, most RNAP organize in clusters that localize at the periphery of the nucleoid.* There is an inverse correlation between the richness of the environment and the number of genes that are expressed genome-wide [54,55]. More precisely, in *E. coli* cells growing in rich media, approximately 90% of the total transcription occurs at operons coding for rRNA (*rrn* operons), although these operons represent only about 1% of the genome. Several transcription cycles may run simultaneously at the same operon, so that clusters of more than 70 RNAP may be involved in the transcription of a single *rrn* operon, while clusters comprising more than 500 RNAP synthesize rRNA from multiple, neighboring operons [6,10]. Recent *in vivo* microscopy experiments indicate that these RNAP clusters are observed almost exclusively at the periphery of the nucleoids [8-10,12,13], as is illustrated in the right column of Fig. 5. This may be rationalized by realizing that large amounts of rRNA are assembled at the same locus over long periods of time, so that a segregation process similar to that described above for functional ribosomes may eventually take place, resulting in a shift the corresponding genes and the bound RNAP enzymes towards the DNA periphery, in spite of the slowness of such large-scale rearrangements. Note that RNAP clusters have also been observed in cells growing in nutrient-poor media, but their sizes remain relatively small (around 35 RNAP) and they gather only a small portion of the total number of RNAP [6]. As for cells growing in rich media, these RNAP clusters, which form in the neighborhood of more strongly expressed genes, are biased towards the nucleoid periphery [10]. Rather interestingly, when transcription (but not fixation of RNAP on promoter sites) is blocked by rifampicin incubation, RNAP clusters no longer localize at the periphery but remain instead buried inside the bulk of the nucleoid [10]. Active synthesis of RNA and the eventual recruitment of ribosomes by nascent mRNA are therefore mandatory for the shift of RNAP clusters and the genes being transcribed to occur. This indicates that entropic forces are not sufficient to provoke the shift and that they must be supplemented with demixing forces resulting from DNA/RNA electrostatic interactions.

*The nucleoid of cells grown in rich media is more compact.* As mentioned above, the nucleoid of cells growing rapidly in rich media is more compact than the nucleoid of cells growing slowly in nutrient-poor media [6-10]. This is clearly related to the different levels of gene expression. Indeed, in cells growing slowly in poor media many genes distributed homogeneously in the nucleoid are expressed at low to moderate levels, which results in the assembly of large negatively charged translation complexes throughout the nucleoid. As a consequence, the nucleoid never reaches the equilibrium fully compacted state. In contrast, the genes most expressed during fast growth in rich media (essentially *rrn* operons) are shifted towards the periphery of the nucleoid and less than 10% of gene expression takes place inside the nucleoid. Therefore, much fewer translation complexes are formed inside the nucleoid, which comes closer to the equilibrium fully compacted state.

*Induction of a gene may drive a shift of its genetic locus toward the periphery of the nucleoid.* It was found that induction of LacY (the membrane protein lactose permease) and TetA (the tetracycline efflux pump) expression in *E. coli* cells grown in poor media leads to a shift of their genetic loci toward the cell membrane (and consequently the periphery of the nucleoid) over time scales as short as 3 minutes [56]. This result confirms that the eventual shift of a gene locus toward the periphery of the nucleoid depends indeed on its expression level. When translation was eliminated by mutating the start codon or incubating the cell with the antibiotics kasugamycin, which inhibits the assembly of functional 70S ribosomes on mRNA [57], the shift was however no longer observed [56]. This indicates that for these two genes both larger transcription rates and assembly of functional ribosomes on nascent mRNA are necessary to induce the shift of the corresponding loci toward the periphery of the nucleoid and the reorganization of the DNA coil. In contrast, no shift of the gene locus was observed upon induction of two cytoplasmic proteins instead of the two membrane proteins LacY and TetA [56]. A tentative explanation for this observation is that translation of the mRNA transcribed from the two former genes would start at the nascent stage while that of the two latter genes would start only when the mRNA are mature transcripts no longer bound to RNAP and DNA, but this hypothesis requires confirmation.

*Cells treated wih rifampicin display decondensed nucleoids*. Recent *in vivo* microscopy observations consistently indicate that *E. coli* cells treated with the rifampicin antibiotics display large decondensed nucleoids that extend almost throughout the cell, as is clearly seen in Fig. 3(A), regardless of the richness of the growth medium [8-10,12,14-16]. This transcription inhibitor acts by blocking the synthesis of the second or third phosphodiester bond of the RNA by the RNAP enzyme [58]. In contrast, RNAP enzymes become totally resistant to rifampicin once they have synthesized a long transcript and entered the elongation phase, so that enzymes already engaged in elongation when the antibiotics is added finish the ongoing transcription (on the time scale of 1 minute) before stalling at the next elongation attempt. Since about 80% of complete mRNA transcripts in *E. coli* cells have half-lives ranging from 3 to 8 minutes [59] and a translation cycle lasts about 1 minute, it may be expected that after 30 minutes almost all cellular mRNA have been degraded by ribonucleases and nearly all ribosomes have consequently converted back to free 30S and 50S subunits. The distribution of ribosomes shown in Figs. 3(B) and 3(C) matches closely that of DNA, which confirms that ribosomes are indeed in the form of free subunits that penetrate and diffuse freely inside the nucleoid and are not able to exert any segregation force on the DNA. Observation that cells treated with rifampicin display decondensed nucleoids is therefore consistent with the hypothesis that nucleoid compaction is driven by DNA/(translation complexes) demixing.

*Nucleoids of cells treated wih chloramphenicol are further compacted.* Contrary to rifampicin, the nucleoid of cells treated with chloramphenicol is much more compact than the nucleoid of untreated cells, regardless of the richness of the growth medium [8-10,13-16], as can be checked in Fig. 4(A). Chloramphenicol is a translation inhibitor, which acts by binding near the peptidyl transferase center of the large ribosomal subunit, thereby preventing further binding of tRNA to the ribosomal A and P sites [60]. Elongation is consequently inhibited and ribosomes remain bound to the mRNA molecule instead of separating into free subunits and being released at the end of the translation run. It may therefore be expected that after a short while all ribosomal subunits are engaged in a stalled translation complex that has been expelled outside from the nucleoid. It can be checked in Figs. 4(B) and Fig. 4(C), which show the ribosome distribution in the cell, that this is indeed the case. In particular, Fig. 4(C) shows the spatial distribution of ribosomes along the short axis of the cell averaged over the portion of the cell occupied by the nucleoid. The peak-to-valley ratio is of the order of 10:1, much larger than for untreated cells, which denotes significantly greater ribosome-nucleoid segregation. All ribosomal subunits being engaged in stalled translation complexes expelled outside the nucleoid, no subunits remain available for the assembly of translation complexes inside the nucleoid. As a consequence, the nucleoid reaches its equilibrium most compact state. Observation that the nucleoid of cells treated with chloramphenicol is very compact is therefore consistent with the hypothesis that nucleoid compaction is driven by DNA/(translation complexes) demixing.

## 5. Crowding may compensate for the short range of electrostatic interactions

As discussed in Section 4 above, recent *in vivo* high-resolution microscopy experiments provide converging indications that segregative phase separation induced by electrostatic repulsion between DNA and RNA (essentially ribosomal RNA) may be the leading force in the formation of the nucleoid. At first sight, this may sound surprising because electrostatic forces are rather short-ranged in the cytosol. The value of the Debye length is not precisely known (it is not even warranted that this is a relevant quantity in so highly concentrated electrolyte solutions), but it is generally estimated to be of the order of 1 nm, and the question arises whether such short range forces may have a significant impact on the organization of the cell. Without pretending to bring a definitive answer to this difficult question, the purpose of this section is to suggest that macromolecular crowding may compensate for the short range of electrostatic forces by constraining macromolecules to remain at short distances from one another and interact in spite of the strong screening.

Confirmation of a possible synergy between crowding and electrostatic forces was gained from simulations performed with the coarse-grained model described in Eqs. (A.11) and (A.12) of Ref. [25]. Briefly, the model consists of a circular chain of 1440 beads enclosed in a confining sphere of radius $R_0 = 120$ nm. Each bead represents 15 base pairs, so that the complete chain represents a DNA sequence with 21600 base pairs at a nucleotide concentration close to the *in vivo* bacterial one. The potential energy function of the DNA chain contains stretching, bending and electrostatic repulsion terms [25]. Upon thermal equilibration, the DNA chain adopts coil conformations that fill the whole confining sphere homogeneously, as shown in Fig. 6(B). $N$ additional beads with radius $b$, which interact with themselves and the DNA chain through repulsive electrostatic terms, are then introduced at homogeneously distributed random positions in the confining sphere (Fig. 6(A)) and the system is allowed to equilibrate again. Standard values were assumed for the parameters of the DNA chain ($h = 1000\, k_B T / l_0^2$ for the stretching force constant, $g = 9.82\, k_B T$ for the bending force constant, $\alpha = 1$ and $\gamma = 0$ in the expression of the interaction energy between two beads that are not nearest-neighbors, which implies pure electrostatic repulsion, and $\zeta = 1000\, k_B T$ for the wall repulsion coefficient, where $T = 298$ K stands for temperature and $l_0 = 5.0$ nm for the equilibrium distance between two successive beads), with the Debye length being set to $r_D = 1.07$ nm. For the $N$ additional beads, the sphere wall repulsion coefficient was also set to $\zeta^{(SP)} = 1000\, k_B T$ and simulations were run with two different values of the bead charge ratio ($\alpha^{(SP)} = 0.8$ and $\alpha^{(SP)} = 1.0$), two different numbers of additional beads ($N = 2000$ and $N = 3000$), and a large number of values of the radius $b$ ranging from 2.8 to 7.5 nm. The dynamics of the complete system composed of the DNA chain and the $N$ additional beads was investigated by integrating numerically overdamped Langevin equations with a time step $\Delta t = 20$ ps, according to Eq. (A.15) of Ref. [25].

Results obtained with $\alpha^{(SP)} = 1.0$ and $N = 2000$ are displayed in Fig. 6(D). This plot shows the evolution of the average radius of gyration $R_g$ of the equilibrated DNA coil as a function of the bead radius $b$. It is seen that the plot can be divided into three regions. For radii smaller than $b \cong 6.0$ nm, demixing between DNA and the additional beads, and consequently compaction of the DNA coil, increases progressively with $b$ but remains quite limited, with a minimum value of $R_g$ around 66 nm. For $b \approx 6.5$ nm, demixing and compaction are instead quite strong, with an average radius of gyration as small as about 50 nm (Fig. 6(C)). Note that this radius corresponds to a volume occupancy ratio $Nb^3 / R_0^3 \approx 0.32$, which is of the same order of magnitude as intracellular values. Finally, for values of $b$ larger than about 6.8 nm, introduction of the $N$ additional beads does not result in any demixing or compaction of the DNA coil ($R_g \approx 83$ nm), for the reason that the system is jammed and cannot reorganize. Simulations performed with { $N = 2000$ and $\alpha^{(SP)} = 0.8$ } and { $N = 3000$ and $\alpha^{(SP)} = 1.0$ } lead to results very similar to those shown in Fig. 6, except that in the latter case strong demixing and compaction are observed for values of $b$ around 5.3 nm instead of 6.5 nm, that is again for a volume occupancy ratio of 0.26. Strikingly, efficient demixing and compaction therefore take place in a rather narrow range of values of $b$ just below the critical value where jamming occurs, thereby emphasizing the importance of crowding to compensate for the short range of electrostatic forces.

In conclusion, these simulations highlight the fact that crowding may indeed work synergetically with electrostatic forces. By constraining macromolecules to remain at short distances from one another and feel electrostatic interactions in spite of the strength of the screening exerted by electrolyte species, crowding may ultimately favor DNA/RNA demixing and nucleoid compaction. Still, one should probably be careful in assigning DNA compaction by other negatively charged species primarily to a "crowding effect", since this latter denomination implicitly refers to entropic depletion forces, which are usually shorter-ranged than screened electrostatic interactions and display opposite trends upon salt concentration variations.

## 6. Conclusion

Recent advances in *in vivo* sub-wavelength resolution microscopy techniques have allowed the observation of the compaction dynamics of bacterial nucleoids at an unprecedented level of detail. In this paper, I have reviewed some of these fascinating new results and argued that they provide converging indications in favor of a model that describes the cytosol as an aqueous electrolyte solution containing several macromolecular species, where demixing occurs between DNA and RNA (essentially rRNA and mRNA involved in translation complexes, but also the large amounts of rRNA synthesized at the *rrn* operons of cells growing in rich media). Several other mechanisms may play a role in the formation of the nucleoid by lowering the free energy which is required to compact the DNA, including neutralization of DNA charges by polycations, fluctuations correlation forces and depletion forces. Crowding probably plays a second, crucial role by working synergetically with electrostatic forces to compensate for their short range and increase strongly the level of DNA/RNA demixing and nucleoid compaction.

The analysis proposed here may be pursued and backed up along several lines. For example, further theoretical and simulation work could aim at investigating more thoroughly the difference between the properties of free 30S and 50S subunits and those of functional translation complexes with respect to the nucleoid compaction dynamics. A first attempt in this direction was described in Ref. [16], but in this work 70S ribosomes were considered only in the form of 13-mer polysomes, so that these simulations essentially pointed out the role of translational entropy rather than that of the $\chi$ parameter. More generally, it would be interesting to gain some insight into how macromolecules that are not expected to take part in the demixing process (ribosomal subunits, but also the NAPs, free mRNA, DNAP, *etc*…) interfere or do not interfere with it. Moreover, as mentioned in the Introduction and illustrated in Fig. 2, the size of the nucleoid fluctuates widely during DNA replication and segregation, as well as during cell division [11,61,62]. It could be interesting to relate the observed variations of the organization of the nucleoid with the biochemical or biophysical mechanisms at play at a given stage along lines similar to those developed here. This is however admittedly a difficult task because replication and segregation are still poorly understood and the mechanisms at work moreover likely depend on the investigated organism [61,62].

**REFERENCES**

[1] Teraoka I. Polymer solutions: An introduction to physical properties. 1st ed. New York: Wiley; 2002.

[2] Vendeville A, Larivière D, Fourmentin E. An inventory of the bacterial macromolecular components and their spatial organization. FEMS Microbiol Rev 2011; 35:395-414.

[3] Bloomfield VA. DNA condensation by multivalent cations. Biopolymers 1998; 44:269-82.

[4] Pelletier J, Halvorsen K, Ha BY, Paparcone R, Sandler SJ, Woldringh CL, Wong WP, Jun S. Physical manipulation of the *Escherichia coli* chromosome reveals its soft nature. Proc Natl Acad Sci USA 2012; 109:E2649-56.

[5] Stracy M, Uphoff S, Garza de Leon F, Kapanidis AN. In vivo single-molecule imaging of bacterial DNA replication, transcription and repair. FEBS Lett 2014; 588:3585-94.

[6]** Endesfelder U, Finan K, Holden SJ, Cook PR, Kapanidis AN, Heilemann M. Multiscale spatial organization of RNA polymerase in *Escherichia coli*. Biophys J 2013; 105:172-81.

[7]** Yazdi NH, Guet CC, Johnson RC, Marko JF. Variation of the folding and dynamics of the *Escherichia coli* chromosome with growth conditions. Mol Microbiol 2012; 86:1318-33.

[8]** Jin DJ, Cagliero C, Martin CM, Izard J, Zhou YN. The dynamic nature and territory of transcriptional machinery in the bacterial chromosome. Front Microbiol 2015; 6:497.

[9]** Jin DJ, Cagliero C, Zhou YN. Role of RNA polymerase and transcription in the organization of the bacterial nucleoid. Chem Rev 2013; 113:8662-82.

[10]** Stracy M, Lesterlin C, Garza de Leon F, Uphoff S, Zawadzki P, Kapanidis AN. Live-cell superresolution microscopy reveals the organization of RNA polymerase in the bacterial nucleoid. P Natl Acad Sci USA 2015; 112:E4390-9.

[11]** Spahn C, Endesfelder U, Heilemann M. Super-resolution imaging of *Escherichia coli* nucleoids reveals highly structured and asymmetric segregation during fast growth. J Struct Biol 2014; 185:243-9.

[12]** Cagliero C, Jin DJ. Dissociation and re-association of RNA polymerase with DNA during osmotic stress response in *Escherichia coli*. Nucleic Acids Res 2013; 41:315-26.

[13]** Cabrera JE, Cagliero C, Quan S, Squires CL, Jin DJ. Active transcription of rRNA operons condenses the nucleoid in *Escherichia coli* : Examining the effect of transcription on nucloid structure in the absence of transertion. J Bacteriol 2009; 191:4180-5.

[14]** Bakshi S, Choi H, Weisshaar JC. The spatial biology of transcription and translation in rapidly growing *Escherichia coli*. Front Microbiol 2015; 6:636.

[15]** Bakshi S, Siryaporn A, Goulian M, Weisshaar JC. Superresolution imaging of ribosomes and RNA polymerase in live *Escherichia coli* cells. Mol Microbiol 2012; 85:21-38.

[16]** Bakshi S, Choi H, Mondal J, Weisshaar JC. Time-dependent effects of transcription- and translation-halting drugs on the spatial distributions of the *Escherichia coli* chromosome and ribosomes. Mol Microbiol 2014; 94:871-887.

[17] Cayley S, Lewis BA, Guttman HJ, Record MT. Characterization of the cytoplasm of *Escherichia coli* K-12 as a function of external osmolarity. J Mol Biol 1991; 222:281-300.

[18] Ishihama Y, Schmidt T, Rappsilber J, Mann M, Hartl FU, Kerner MJ, Frishman D. Protein abundance profiling of the *Escherichia coli* cytosol. BMC Genomics 2008; 9:102.

[19] Roe AJ, McLaggan D, Davidson I, O'Byrne C, Booth IR. Perturbation of anion balance during inhibition of growth of *Escherichia coli* by weak acids. J Bacteriol 1998; 180:767-72.

[20] Bennett BD, Kimball EH, Gao M, Osterhout R, Van Dien SJ, Rabinowitz JD. Absolute metabolite concentrations and implied enzyme active site occupancy in *Escherichia coli*. Nat Chem Biol 2009; 5:593-9.

[21] Shabala L, Bowman J, Brown J, Ross T, McMeekin T, Shabala S. Ion transport and osmotic adjustment in *Escherichia coli* in response to ionic and non-ionic osmotica. Environ Microbiol 2009; 11:137-48.

[22] Moncany MLJ, Kellenberger E. High magnesium content of *Escherichia coli* B. Experientia 1981; 37:846-7.

[23] Capp MW, Cayley DS, Zhang W, Guttman HJ, Melcher SE, Saecker RM, Anderson CF, Record MT. Compensating effects of opposing changes in putrescine(2+) and $K^+$ concentrations on *lac* repressor-*lac* operator binding: *in vitro* thermodynamic analysis and *in vivo* relevance. J Mol Biol 1996; 258:25-36.

[24] Manning GS. Limiting Laws and Counterion Condensation in Polyelectrolyte Solutions I. Colligative Properties. J Chem Phys 1969; 51:924-33.

[25] Joyeux M. Compaction of bacterial genomic DNA : clarifying the concepts. J Phys Condens Matter 2015; 27:383001.

[26] Wilson RW, Bloomfield VA. Counterion-induced condensation of deoxyribonucleic acid. A light-scattering study. Biochemistry 1979; 18:2192-6.

[27] Gosule LC, Schellman JA. Compact form of DNA induced by spermidine. Nature 1976; 259:333-5.

[28] Marquet R, Houssier C. Thermodynamics of cation-induced DNA condensation. J Biomol Struct Dyn 1991; 9:159-67.

[29] Asakura S, Oosawa F. On interaction between two bodies immersed in a solution of macromolecules. J Chem Phys 1954; 22:1555-6.

[30] Lerman LS. A transition to a compact form of DNA in polymer solutions. Proc Natl Acad Sci USA 1971; 68:1886-90.

[31] Stuger R, Woldringh CL, van der Weijden CC, Vischer NOE, Bakker BM, van Spanning RJM, Snoep JL, Westerhoff HV. DNA supercoiling by gyrase is linked to nucleoid compaction. Mol Biol Rep 2002; 29:79-82.

[32] Cunha S, Woldringh CL, Odijk T. Polymer-mediated compaction and internal dynamics of isolated *Escherichia coli* nucleoids. J Struct Biol 2001; 136:53-66.

[33] Azam TA, Iwata A, Nishimura A, Ueda S, Ishihama A. Growth phase-dependent variation in protein composition of the *Escherichia coli* nucleoid. J Bacteriol 1999; 181:6361-70.

[34] Sato YT, Watanabe S, Kenmotsu T, Ichikawa M, Yoshikawa Y, Teramoto J, Imanaka T, Ishihama A, Yoshikawa K. Structural change of DNA induced by nucleoid proteins: Growth phase-specific Fis and stationary phase-specific Dps. Biophys J 2013; 105:1037-44.

[35] Lin J, Chen H, Dröge P, Yan J. Physical organization of DNA by multiple non-specific DNA-binding modes of Integration Host Factor (IHF). PloS One 2012; 7:e49885.

[36] Sagi D, Friedman N, Vorgias C, Oppenheim AB, Stavans J. Modulation of DNA conformations through the formation of alternative high-order HU-DNA complexes. J Mol Biol 2004; 341:419-28.

[37] Beloin C, Jeusset J, Révet B, Mirambeau G, Le Hégarat F, Le Cam E. Contribution of DNA conformation and topology in right-handed DNA wrapping by the *Bacillus subtilis* LrpC protein. J Biol Chem 2003; 278:5333-42.

[38] Joyeux M, Vreede J. A model of H-NS mediated compaction of bacterial DNA. Biophys J 2013; 104:1615-22.

[39] Dame RT, Wyman C, Goosen N. H-NS mediated compaction of DNA visualised by atomic force microscopy. Nucleic Acids Res 2000; 28:3504-10.

[40] Joyeux M. Equilibration of complexes of DNA and H-NS proteins on charged surfaces: A coarse-grained model point of view. J Chem Phys 2014; 141:115102.

[41] Spurio R, Dürrenberger M, Falconi M, La Teana A, Pon CL, Gualerzi CO. Lethal overproduction of the *Escherichia coli* nucleoid protein H-NS: ultramicroscopic and molecular autopsy. Mol Gen Genet 1992; 231:201-11.

[42] Paull TT, Johnson RC. DNA looping by *Saccharomyces cerevisiae* high mobility group proteins NHP6A/B. J Biol Chem 1995; 270:8744-54.

[43] Espéli O, Mercier R, Boccard F. DNA dynamics vary according to macrodomain topography in the *E. coli* chromosome. Mol Microbiol 2008; 68:1418-27.

[44] Mercier R, Petit MA, Schbath S, Robin S, El Karoui M, Boccard F, Espéli O. The MatP/*matS* site-specific system organizes the terminus region of the *E. coli* chromosome into a macrodomain. Cell 2008; 135:475-85.

[45] Hsu CC, Prausnitz JM. Thermodynamics of polymer compatibility in ternary systems. Macromolecules 1974; 7:320-4.

[46] Laemmli UK. Characterization of DNA condensates induced by poly(ethylene oxide) and polylysine. Proc Natl Acad Sci USA 1975; 72:4288-92.

[47] Akitaya T, Seno A, Nakai T, Hazemoto N, Murata S, Yoshikawa K. Biomacromolecules 2007; 8:273-8.

[48]** Krotova MK, Vasilevskaya VV, Makita N, Yoshikawa K, Khokhlov AR. DNA compaction in a crowded environment with negatively charged proteins. Phys Rev Lett 2010; 105:128302.

[49]** Yoshikawa K, Hirota S, Makita N, Yoshikawa Y. Compaction of DNA induced by like-charge protein: opposite salt-effect against the polymer-salt-induced condensation with neutral polymer. J Phys Chem Lett 2010; 1:1763-6.

[50] Wagh J, Patel KJ, Soni P, Desai K, Upadhyay P, Soni HP. Transfecting pDNA to *E. coli* DH5α using bovine serum albumin nanoparticles as a delivery vehicle. Luminescence 2015; 30:583-91.

[51]** Zinchenko A, Tsumoto K, Murata S, Yoshikawa K. Crowding by anionic nanoparticles causes DNA double-strand instability and compaction. J Phys Chem B 2014; 118:1256-62.

[52]** Sanamrad A, Persson F, Lundius EG, Fange D, Gynna AH, Elf J. Single-particle tracking reveals that free ribosomal subunits are not excluded from the *Escherichia coli* nucleoid. P Natl Acad Sci USA 2014; 111:11413-8.

[53] Proshkin S, Rahmouni AR, Mironov A, Nudler E. Cooperation between translating ribosomes and RNA polymerase in transcription elongation. Science 2010; 328:504-8.

[54] Liu M, Durfee T, Cabrera JE, Zhao K, Jin DJ, Blattner FR. Global transcriptional programs reveal a carbon source foraging strategy by *Escherichia coli*. J Biol Chem 2005; 280:15921-7.

[55] Tao H, Bausch C, Richmond C, Blattner FR, Conway T. Functional genomics: Expression analysis of *Escherichia coli* growing on minimal and rich media. J Bacteriol 1999; 181:6425-40.

[56]** Libby EA, Roggiani M, Goulian M. Membrane protein expression triggers chromosomal locus repositioning in bacteria. P Natl Acad Sci USA 2012; 109:7445-50.

[57] Mankin A. Antibiotic blocks mRNA path on the ribosome. Nat Struct Mol Biol 2006; 113:858-60.

[58] Campbell EA, Korzheva N, Mustaev A, Murakami K, Nair S, Goldfarb A, Darst SA. Structural mechanism for rifampicin inhibition of bacterial RNA polymerase. Cell 2001; 104:901-12.

[59] Bernstein JA, Khodursky AB, Lin PH, Lin-Chao S, Cohen SN. Global analysis of mRNA decay and abundance in *Escherichia coli* at single-gene resolution using two-color fluorescent DNA microarrays. Proc Natl Acad Sci USA 2002; 99:9697-702.

[60] Hansen JL, Moore PB, Steitz TA. Structures of five antibiotics bound at the peptidyl transferase center of the large ribosomal subunit. J Mol Biol 2003; 330:1061-75.

[61] Fisher JK, Bourniquel A, Witz G, Weiner B, Prentiss M, Kleckner N. Four-dimensional imaging of *E. coli* nucleoid organization and dynamics in living cells. Cell 2013; 153:882-95.

[62] Berlatzky IA, Rouvinski A, Ben-Yehuda S. Spatial organization of a replicating bacterial chromosome. P Natl Acad Sci USA 2008; 105:14136-40.

**Figure 1** : Influence of the richness of the growth medium on the nucleoid of *E. coli* cells.

(**A**) Differential interference contrast (DIC) images of *E. coli* cells growing and dividing in grooves under rich LB agarose pad and dividing approximately every 30 min at 30°C. (**B**) Fluorescence images of the chromosomes of the same cells as in (A). (**C**) DIC images of cells growing and dividing in grooves under minimal AB glucose-acetate-agarose pad and dividing approximately every 110 min at 30°C. (**D**) Fluorescence images of the chromosomes of the same cells as in (C). The fluorescence images were obtained by fusing the Green Fluorescent Protein (GFP) to the major nucleoid associated protein Fis, which binds DNA tightly and non-specifically. Scale bars: 2 μm. Adapted from Figs. 1 and 3 of Ref. [7], with permission from John Wiley and Sons (Copyright 2015).

**Figure 2** : The organisation of *E. coli* nucleoids at different cell cycle stages for cells grown in a LB rich medium at 37°C.

The arrow and the associated numbers indicate the bacterial cell length (in μm), which is directly related to the stage of the cell cycle. For each cell, a diffraction-limited image (upper left), a bright-light image (upper right), a *d*STORM image (middle) and a cartoon depicting chromosomal organization (below) are shown. The dSTORM images were obtained by allowing the cells to incorporate the thymidine analogue EdU in the genomic DNA and attaching a photoswitchable organic dye to this analogue by click chemistry. Scale bars: 1 μm. Reprinted from Fig. 1 of Ref. [11], with permission from Elsevier (Copyright 2014).

**Figure 3** : Effects of treatment with rifampicin of *E. coli* cells growing in EZRDM at 30°C.

(**A**) Wide field ribosome and DNA spatial distributions 30 min after rifampicin addition. Scale bar: 1 μm. (**B**) Super-resolution image of ribosome distribution. (**C**) Wide field intensity distributions for ribosome and DNA along the short axis cell *y*. Images were obtained by staining DNA with DRAQ5 and labeling small ribosomal subunits with YFP. Adapted from Fig. 7 of Ref. [15] by permission of John Wiley and Sons (Copyright 2012).

**Figure 4** : Effects of treatment with chloramphenicol of *E. coli* cells growing in EZRDM at 30°C.

(**A**) Wide field ribosome and DNA spatial distributions 30 min after chloramphenicol addition. Scale bar: 1 μm. (**B**) Super-resolution image of ribosome distribution. (**C**) Ribosome distribution along the short axis cell *y*, averaged over the 1 μm swath through the cell centre (*i.e.* between the two gray ticks in vignette B). Images were obtained by staining DNA with DRAQ5 and labeling small ribosomal subunits with YFP. Adapted from Fig. 8 of Ref. [15] by permission of John Wiley and Sons (Copyright 2012).

**Figure 5** : Distribution of RNAP enzymes in *E. coli* cells.

3D surface renderings of RNAP (red) and DNA (blue) distributions in example cells grown in minimal (Left) and rich (Right) media. Taken from Fig. 5 of Ref. [10].

**Figure 6** : Simulation results highlighting the synergy between crowding and electrostatic interactions.

(**A**) Typical conformation of the equilibrated DNA chain (small red beads) enclosed in the confining sphere and the additional 2000 negatively charged beads (large blue beads), which have just been introduced at random positions. Radius of the additional beads is *b*=6.5 nm. (**B**) Same snapshot as (A), with the additional beads having been removed and only the links between the centers of the DNA beads being shown. (**C**) Same as (B) after the system has been allowed to equilibrate. (**D**) Plot the average radius of gyration $R_g$ of the equilibrated DNA coil as a function of the radius *b* of the additional beads.

**FIGURE 1**

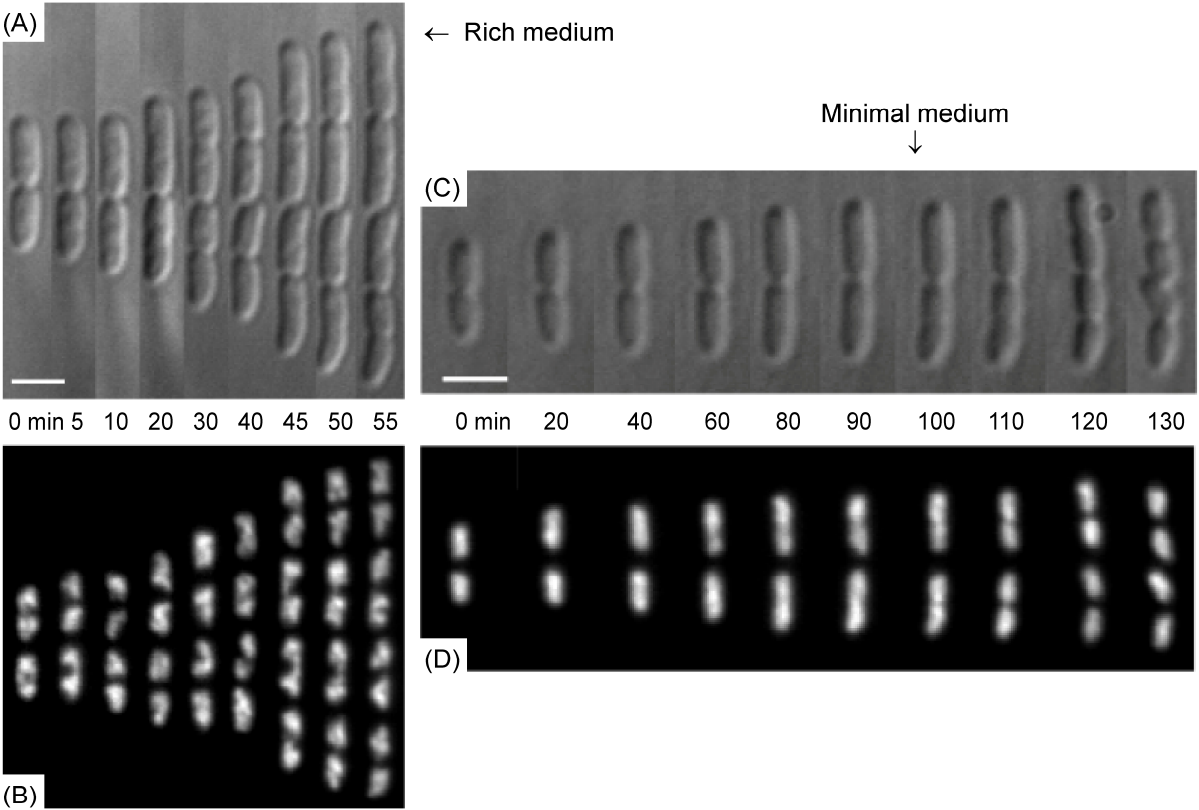

**FIGURE 2**

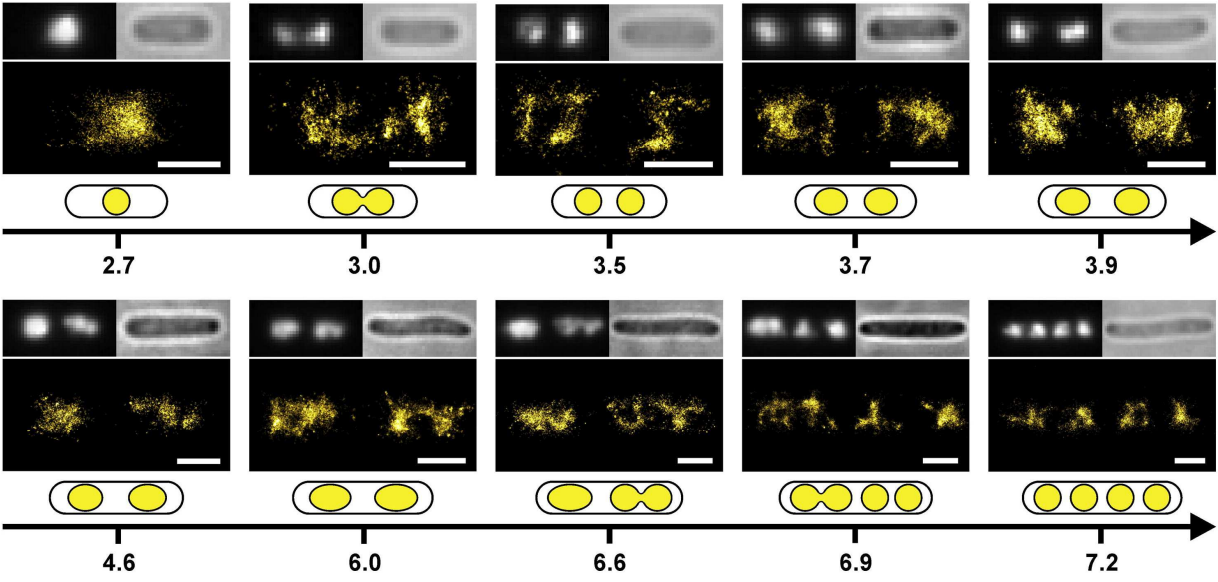

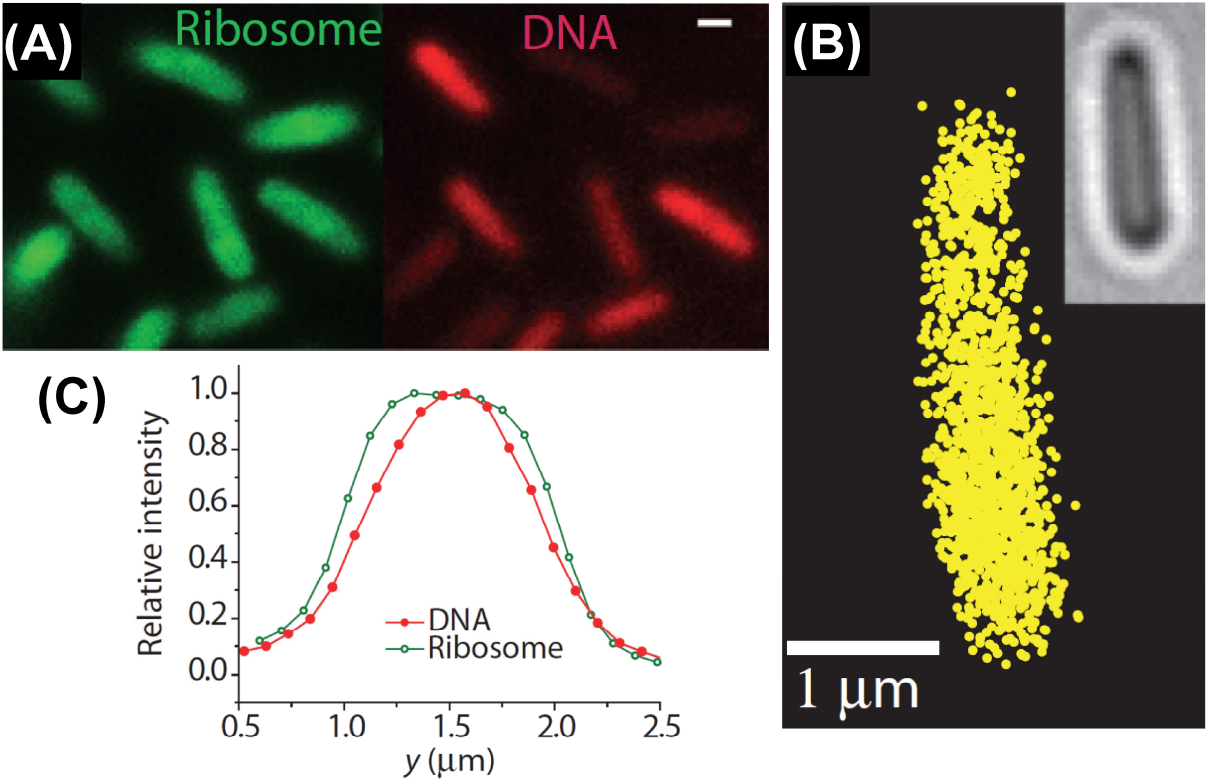

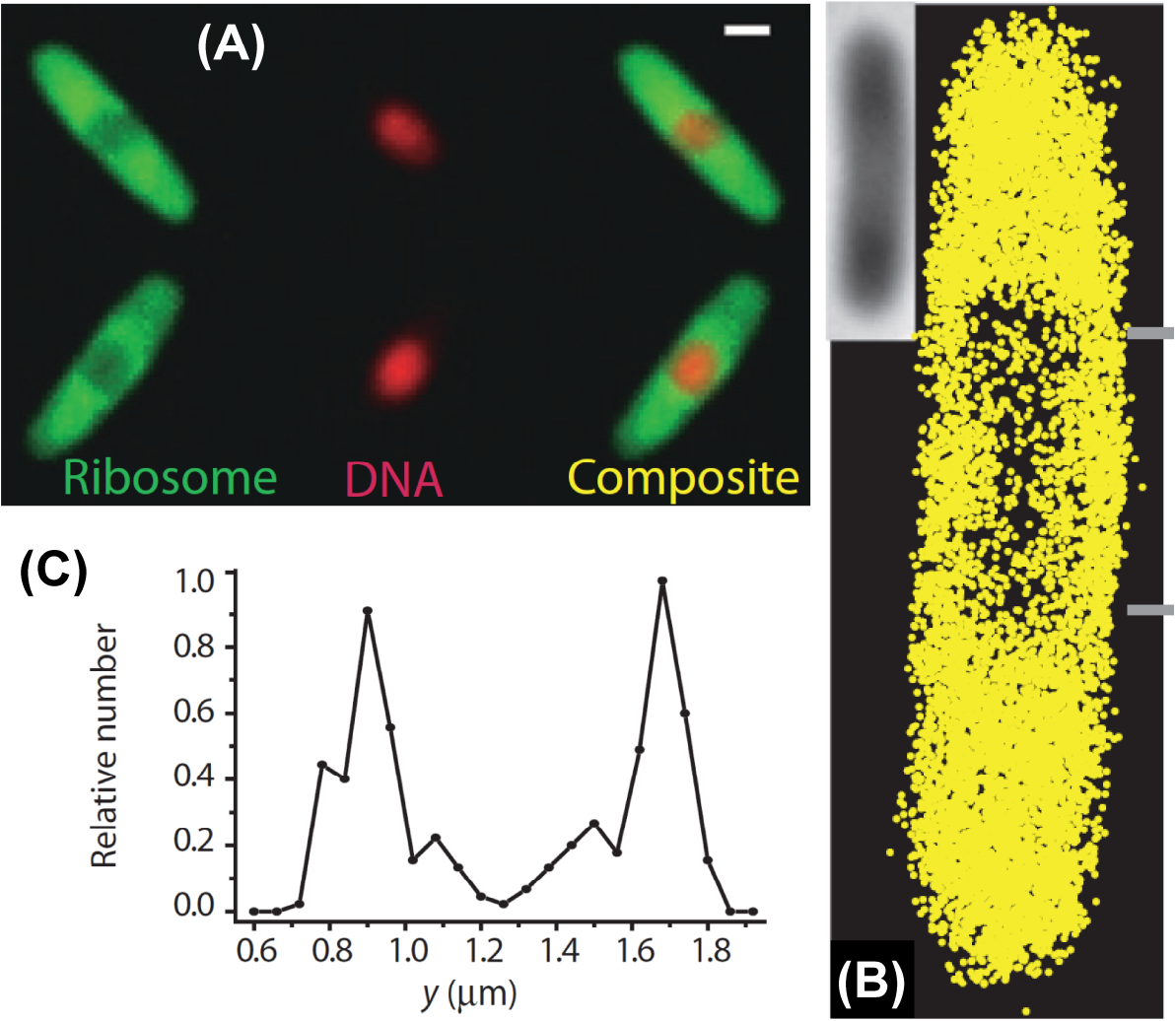

**FIGURE 5**

<u>Minimal medium</u>                    <u>Rich medium</u>

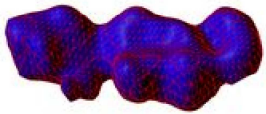    DNA    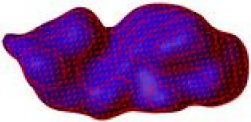    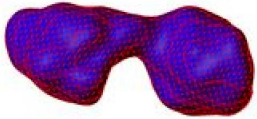

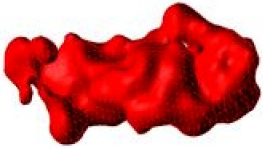    RNAP    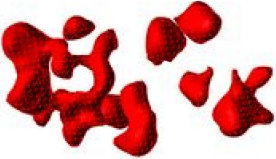    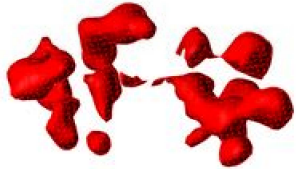

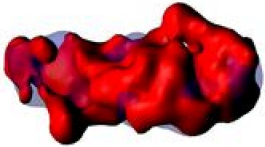    combined    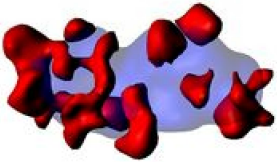    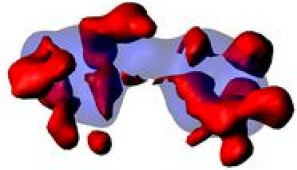

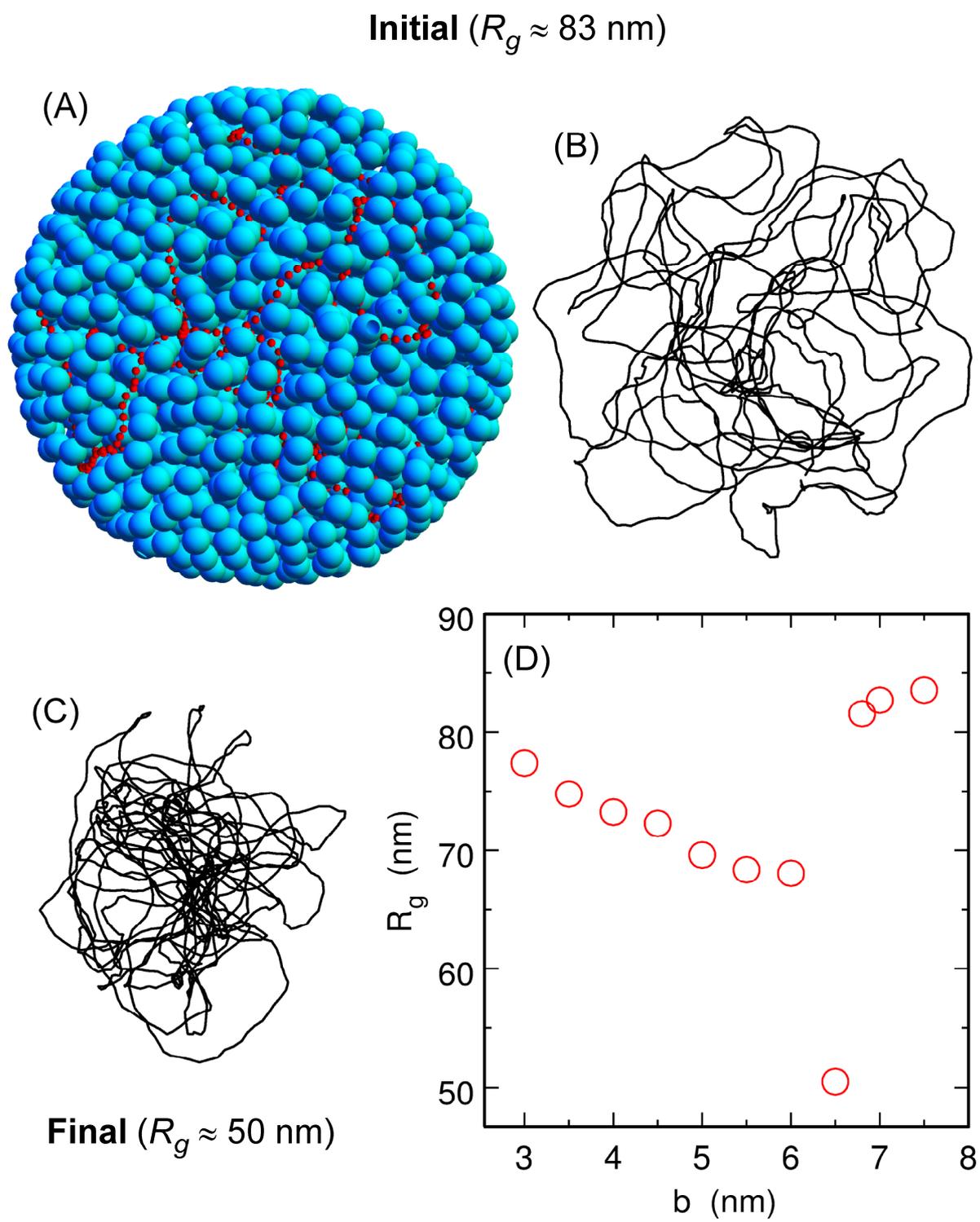

**Initial** ($R_g \approx 83$ nm)

(A)

(B)

(C)

**Final** ($R_g \approx 50$ nm)

(D)